\documentclass[a4paper,11pt]{article}
\usepackage{jcappub} 
\usepackage{lineno}

\usepackage{bm}

\arxivnumber{1234.56789} 
\title{\boldmath Limited Impact of CMB B-mode Polarization on Constraints on Cosmic Topology}

\author[a]{Ryota Himeno}
\author[b,c]{Atsushi J. Nishizawa} 
\author[a,c,d]{Kiyotomo Ichiki} 
\affiliation[a]{Graduate School of Science, Nagoya University, Furocho, Chikusa-ku, Nagoya, Aichi, 464-8602, Japan}
\affiliation[b]{Gifu Shotoku Gakuen University, 1-1, Takakuwanishi, Yanaizucho, Gifu, 501-6194, Japan}
\affiliation[c]{Kobayashi-Maskawa Institute for the Origin of Particles and the Universe, Nagoya University, Furocho, Chikusa-ku, Nagoya, Aichi, 464-8602, Japan}
\affiliation[d]{Institute for Advanced Research, Nagoya University, Furocho, Chikusa-ku, Nagoya, Aichi 464-8602, Japan}

\emailAdd{himeno.ryota.w8@s.mail.nagoya-u.ac.jp, 
atsushi.nisizawa@gifu.shotoku.ac.jp,
ichiki.kiyotomo.a9@f.mail.nagoya-u.ac.jp}

\abstract{
The global spatial structure of the universe is a fundamental question in cosmology. Although the standard model usually assumes a trivial topology, a multiply connected universe can break statistical isotropy and generate off-diagonal covariance between the spherical-harmonic coefficients of cosmic microwave background fluctuations.
In this work, we investigate whether CMB B-mode polarization improves constraints on the cubic three-torus, or $E_1$ topology. We construct the full harmonic-space covariance matrices of the temperature, E-mode, and primordial B-mode anisotropies, including correlations between different $(\ell,m)$ modes, and evaluate Gaussian likelihoods using random realizations generated from a statistically isotropic $\Lambda$CDM model with trivial topology. To isolate the intrinsic topological information carried by each CMB observable, we consider the ideal cosmic-variance-limited case, neglecting instrumental noise and gravitational lensing. We find that adding primordial B-mode polarization yields only a marginal improvement in the ability to distinguish the $E_1$ topology from the trivial topology, even under these ideal conditions. The different constraining powers of the temperature, E-mode, and B-mode anisotropies can be understood through the combined effects of their transfer-function responses and the topology-induced off-diagonal correlations in harmonic space.
}

\begin{document}
\maketitle
\flushbottom
\section{Introduction} 
The Friedmann–Lemaître–Robertson–Walker (FLRW) metric describes a homogeneous and isotropic universe whose spatial sections have constant curvature~\cite{1997gr.qc....12019C}. 
Depending on the sign of the curvature parameter, the corresponding simply connected covering space is Euclidean space $\mathbb{R}^3$, the three-sphere $S^3$, or hyperbolic space $H^3$.
However, the local curvature does not uniquely determine the global topology of space. 
A spatial manifold with covering space $X \in \{\mathbb{R}^3, S^3, H^3\}$ may instead be written as a quotient $X/\Gamma$, where $\Gamma$ is a discrete group of fixed-point-free isometries~\cite{2002PhR...365..251L}. 
Thus, the Einstein equations determine the local geometry of the universe but do not, by themselves, specify whether its spatial sections are simply or multiply connected~\cite{1971GReGr...2....7E}.

The global topology of the universe has attracted interest in both mathematics and cosmology. 
In mathematics, Thurston's geometrization conjecture provided a unified framework for understanding three-dimensional manifolds~\cite{1982BAMS....6..357T}, and its subsequent proof by Perelman also resolved the Poincaré conjecture~\cite{2002math.....11159P,2003math......3109P}. 
In cosmology, the Circles-in-the-Sky (CitS) method introduced a direct observational test of non-trivial topology by searching for matched temperature patterns in the cosmic microwave background~\cite{1998CQGra..15.2657C,2004PhRvL..92t1302C}.

The standard spatially flat $\Lambda$CDM model is usually formulated on the simply connected Euclidean space $\mathbb{R}^3$. 
However, vanishing spatial curvature does not uniquely determine the global topology of the universe. 
In the standard classification of Euclidean three-manifolds, $\mathbb{R}^3$ is only one of 18 possible topologies, the remaining 17 being non-trivial~\cite{Wolf1967,1995PhR...254..135L}. 
In this work, we focus on the simplest compact case, the three-torus $T^3$ ($E_1 \cong S^1 \times S^1 \times S^1 \cong \mathbb{R}^3 / \mathbb{Z}^3$), obtained by identifying opposite faces of a fundamental domain through translations. 
We further restrict the fundamental domain to a cube characterized by a single side length $L$.

A non-trivial topology can modify the global symmetries of space even when its local geometry remains flat. 
In the cubic three-torus considered in this work, spatial homogeneity is preserved, whereas continuous rotational invariance is reduced to the discrete symmetries of the fundamental cube~\cite{1995PhR...254..135L,2016Univ....2....1L}. 
Consequently, the CMB fluctuations are not statistically isotropic: spherical-harmonic coefficients with different $(\ell,m)$ modes can be correlated, giving rise to non-vanishing off-diagonal elements in their covariance matrix. 
These correlations provide the primary observational signature of cosmic topology investigated in this work. 

The observability of a non-trivial cosmic topology depends on its evolution during inflation. 
A sufficiently long period of inflation would stretch any pre-existing fundamental domain beyond the present observable universe, making its topological signatures difficult to detect~\cite{2004JCAP...10..004L}. 
Nevertheless, compact spatial topologies have been considered in quantum-cosmological scenarios for the creation of the Universe~\cite{1984PAZh...10..323Z,1983PhRvD..28.2960H,2004JCAP...10..004L}. 
In addition, some string-inspired inflationary models impose geometric constraints on the inflaton field range and, hence, on the possible duration of inflation~\cite{2007PhRvD..75l3508B}. 
These considerations leave open the possibility that a primordial topology could remain observable on cosmological scales.

Motivated by these theoretical considerations, cosmic topology has been explored observationally using both the three-dimensional distribution of astronomical objects and full-sky CMB data from missions such as WMAP and Planck. One class of methods searches for repeated images, or topological clones, of quasars and galaxies produced by the global identifications of space~\cite{1996A&A...313..339L,1998tx19.confE.510F}.
A second class uses the temperature and polarization anisotropies of the CMB~\cite{1998CQGra..15.2657C,2004PhRvL..92t1302C,2016A&A...594A..18P}. The Circles-in-the-Sky method, for example, exploits the fact that if the fundamental domain is sufficiently small, the last-scattering surface intersects one or more of its topological images. This produces pairs of circles on the sky with correlated temperature patterns.
Searches using WMAP and Planck data have found no statistically significant matched-circle pairs and have consequently placed lower bounds on the size of the fundamental domain~\cite{2004PhRvL..92t1302C,2016A&A...594A..18P,2023JCAP...01..030P}.
Complementary approaches use the full covariance of the spherical-harmonic coefficients, including correlations between different $(\ell,m)$ modes. More recently, machine-learning techniques have been applied directly to cosmic-topology classification \cite{2024JCAP...09..057T}, building on spherical neural-network methods developed for CMB and cosmological data~\cite{2019A&A...628A.129K,2019A&C....27..130P}.

Non-trivial topology has also been proposed as a possible explanation for several large-angle features of the observed CMB. These include the suppression of power and correlations at low multipoles~\cite{2016CQGra..33r4001S,2004CQGra..21.4901A,2003Natur.425..593L}, as well as the apparent alignment of low-$\ell$ multipoles, often referred to as the ``Axis of Evil''~\cite{2005PhRvL..95g1301L,2007MNRAS.378..153L}. Although these anomalies are not unique signatures of topology and their statistical interpretation remains debated, they motivate testing departures from statistical isotropy using information beyond the conventional angular power spectrum.

Beyond the regime accessible to matched-circle searches, cosmic topology may still be probed through statistical correlations among CMB multipoles~\cite{2024PhRvL.132q1501A,2015PhRvD..92d3003F}. 
Previous analyses have shown that temperature and E-mode polarization can retain sensitivity to topological signatures even for fundamental domains larger than the last-scattering surface~\cite{2024JCAP...03..036E,2015PhRvD..92d3003F}.
Planck has also applied full-covariance searches using temperature and polarization data to constrain non-trivial topology~\cite{2016A&A...594A..18P}.
Recent work by the COMPACT Collaboration has extended the calculation of topology-induced covariance matrices to tensor perturbations, including $T$-, $E$-, and $B$-mode polarization, and has suggested that tensor modes may provide complementary information at higher multipoles~\cite{2025JCAP...08..015S}. 
Primordial B-mode polarization is particularly interesting in this context because, at linear order and in the absence of gravitational lensing, it is sourced exclusively by tensor perturbations, whereas the temperature and E-mode anisotropies are dominated by scalar perturbations~\cite{1997PhRvD..55.7368K,1997PhRvL..78.2054S,1997NewA....2..323H,2016ARA&A..54..227K}. 
If primordial scalar and tensor perturbations are statistically uncorrelated, as assumed in this work, the primordial B-mode signal is independent of the scalar-induced temperature and E-mode anisotropies. 
It therefore provides a complementary channel for probing topology through the tensor sector.

The observational relevance of such an analysis is increasing with the development of future high-sensitivity CMB polarization missions, most notably LiteBIRD~\cite{2023PTEP.2023d2F01L,2020SPIE11443E..2FH}. LiteBIRD is designed to perform a full-sky measurement of CMB polarization and to constrain the tensor-to-scalar ratio with a target uncertainty of $\sigma (r)<0.001$~\cite{2023PTEP.2023d2F01L}. Although the primordial B-mode signal is expected to be weak and current observations provide only upper limits on its amplitude~\cite{2021PhRvL.127o1301B,2020A&A...641A..10P}, future measurements motivate a quantitative assessment of whether B-mode polarization contains useful additional information about cosmic topology.

This work quantifies how much primordial B-mode polarization improves the detectability of non-trivial cosmic topology. We assume a spatially flat $\Lambda$CDM model with trivial topology as the fiducial universe~\cite{2020A&A...641A...6P} and compare it with the cubic three-torus, or $E_1$, topology. Using the full harmonic-space covariance matrices of the CMB temperature, E-mode, and B-mode anisotropies, we generate mock realizations and evaluate the relative likelihood of the topology models through an effective $\Delta \chi^2$. As a first step, we restrict our analysis to the $E_1$ topology, which provides the simplest compact Euclidean model~\cite{1993JETPL..57..622S}.

This paper is organized as follows. In Section~2, we derive the CMB
covariance matrices for the $E_1$ topology and describe their numerical
implementation. In Section~3, we introduce the Gaussian likelihood and
mock-data generation procedure, and present illustrative CMB maps,
covariance matrices, and the likelihood-ratio analysis. In Section~4, we
discuss the physical origin of the limited B-mode contribution, compare our
results with previous studies, and outline future prospects. Section~5
summarizes our main conclusions. Appendix~A provides a derivation of the
real representation of the harmonic-space Gaussian likelihood.
As discussed in Ref.~\cite{2014A&A...571A..26P}, the ability to detect cosmic topology is insensitive to the cosmological parameters assumed, but for clarity, we assume the best-fit cosmological parameters consistent with Planck 2018 \cite{2020A&A...641A...6P}.

\section{CMB anisotropies in the $E_{1}$ topology}
\label{sec:theory}
\subsection{Harmonic-space covariance in the $E_1$ topology}
\label{subsec:analytic}
In this section, we review the harmonic-space covariance of the CMB anisotropies in the cubic $E_1$ topology, following~\cite{2024JCAP...09..057T}, and extend the formalism to primordial B-mode polarization. In a spatially flat $\Lambda$CDM model with trivial topology, statistical isotropy implies
\begin{equation}
    \left\langle a_{\ell m} a_{\ell' m'}^* \right\rangle
    = C_\ell \delta_{\ell \ell'} \delta_{m m'}~,
\end{equation}
so that all $m$ modes at fixed $\ell$ are statistically equivalent, where $a_{\ell m}$ is the multipole expansion coefficient of CMB anisotropies. In the $E_1$ topology, however, the global spatial identifications reduce continuous rotational symmetry to the discrete symmetries of the fundamental domain. The harmonic-space covariance is therefore generally non-diagonal in $(\ell,m)$, and the full covariance matrix must be retained rather than compressing the information into the conventional angular power spectrum $C_\ell$.
In a usual flat $\Lambda$CDM Universe with trivial topology, we can write the temperature fluctuations in harmonic space as
\begin{equation}
    a_{\ell m} = 4\pi \int \frac{d^3k}{(2\pi)^3} \delta_{\mathcal{R}}(\mathbf{k}) \Delta_\ell(k) \left[ i^\ell Y_{\ell m}^*(\hat{\mathbf{k}}) e^{-i\mathbf{k} \cdot \mathbf{x}_0} \right],
    \label{Eq:prealm}
\end{equation}
where $\delta_{\cal R}$ and $\Delta_\ell(k)$ are the initial curvature perturbation and the temperature transfer function, respectively,
and $\mathbf{x}_0$ represents the position of the observer. 

In the $E_1$ topology, the periodic boundary conditions restrict the allowed wavevectors to $\bm{k}_n =  2\pi \bm{n}/L$, where $L$ denotes the side length of the cubic fundamental domain, with $\bm{n}=(n_1, n_2, n_3)$ and $k_n=|\bm{k}_n|$, where each $n_i$ is an integer. Accordingly, the Fourier integral is replaced by a discrete sum
\begin{equation}
    \int \frac{d^3k}{(2\pi)^3} \to \frac{1}{L^3} \sum_{\mathbf{n}}~,
\end{equation}
and Eq.~(\ref{Eq:prealm}) becomes
\begin{equation}
    a_{\ell m} = \frac{4\pi}{L^3} \sum_{\mathbf{n} \in N} \delta_{\mathcal{R}}(\mathbf{k}_{\mathbf{n}}) \Delta_\ell(k_n) \left[ i^\ell Y_{\ell m}^*(\hat{\mathbf{k}}_{\mathbf{n}}) e^{-i\mathbf{k}_{\mathbf{n}} \cdot \mathbf{x}_0} \right]~.
    \label{Eq:alm}
\end{equation}
For later convenience, we define the geometric factor for temperature fluctuations as
\begin{equation}
    \xi_{k_n \ell m}^{T; \hat{\mathbf{k}}_{\mathbf{n}}} \equiv e^{-i\mathbf{k}_{\mathbf{n}} \cdot \mathbf{x}_0} i^\ell Y_{\ell m}^*(\hat{\mathbf{k}}_{\mathbf{n}})~,
\label{scalar_xi}
\end{equation} 
and the corresponding factor for E- and B-mode fluctuations as 
\begin{equation}
    \xi_{k_n \ell m}^{[E,B];\hat{\mathbf{k}}_{\mathbf{n}}} \equiv \sqrt{\frac{(\ell+2)!}{(\ell-2)!}} e^{-i\mathbf{k}_{\mathbf{n}} \cdot \mathbf{x}_0} i^\ell \, Y_{\ell m}^*(\hat{\mathbf{k}}_{\mathbf{n}})~.
\label{tensor_xi}
\end{equation}
Using Eq.~(\ref{scalar_xi}), the scalar-induced temperature anisotropy Eq.~(\ref{Eq:alm}) can be written compactly as

\begin{equation}
    a_{\ell m}
    =
    \frac{4\pi}{L^3}
    \sum_{\mathbf{n} \in N}
    \delta_{\mathcal{R}}(\mathbf{k}_{\mathbf{n}})
    \Delta_\ell(k_n)
    \xi_{k_n \ell m}^{T; \hat{\mathbf{k}}_{\mathbf{n}}}~.
\end{equation}

Taking the ensemble average of the square of $a_{\ell m}$, we obtain the covariance in the $E_1$ topology model as 
\begin{equation}
    \label{eq:clmlm}
    C_{\ell m \ell' m'} = \frac{(4\pi)^2}{L^3} \sum_{\mathbf{n} \in N} \Delta_\ell(k_n) \Delta_{\ell'}^*(k_n) \frac{2\pi^2 \mathcal{P}_{\mathcal{R}}(k_n)}{k_n^3} \xi_{k_n \ell m}^{T; \hat{\mathbf{k}}_n} \xi_{k_n \ell' m'}^{T;\hat{\mathbf{k}}_n *},
\end{equation}
where $\mathcal{P}_{\mathcal{R}}(k)$ denotes the dimensionless primordial curvature power spectrum,

\begin{equation}
\left\langle
\delta_{\mathcal{R}}(\mathbf{k}_{\mathbf{n}})
\delta_{\mathcal{R}}^*(\mathbf{k}_{\mathbf{n}'})
\right\rangle
=
L^3
\frac{2\pi^2}{k_n^3}
\mathcal{P}_{\mathcal{R}}(k_n)
\delta^K_{\mathbf{n}\mathbf{n}'}~.
\end{equation}

In this work, we extend the covariance formalism to include primordial B-mode polarization. 
To assess the additional information provided by polarization, we consider three data combinations: temperature alone (T), temperature and E-mode polarization (T+E), and temperature, E-mode, and B-mode polarization (T+E+B). 
We assume the T- and E-mode anisotropies are sourced solely from scalar perturbations and the B-mode anisotropies solely from tensor perturbations. 
We further assume that the primordial scalar and tensor perturbations are statistically uncorrelated. 
Under these assumptions, the TB and EB cross-covariances vanish, and the covariance matrix of the concatenated data vector 
\begin{equation}
    \bm{a} = \{a_{\ell m}^T, a_{\ell m}^E, a_{\ell m}^B \}
\end{equation}
takes the block-diagonal form:
\begin{equation}
\mathbf{C}_{\text{model}} = 
\begin{pmatrix}
\mathbf{C}^{TT} & \mathbf{C}^{TE} & \mathbf{0} \\
\mathbf{C}^{ET} & \mathbf{C}^{EE} & \mathbf{0} \\
\mathbf{0} & \mathbf{0} & \mathbf{C}^{BB}
\end{pmatrix}~,
\hspace*{1cm}
\mathbf{C}^{ET} = \left(\mathbf{C}^{TE} \right)^\dagger~.
\end{equation}
We emphasize that this block structure follows from the approximations adopted here, rather than from parity symmetry alone. 

In a statistically anisotropic universe, parity symmetry requires the TB and EB covariances to vanish only for even $\ell+\ell'$, while correlations with odd $\ell+\ell'$ can remain nonzero \cite{1997PhRvD..55.1830Z}.

The individual elements $C^{XY}_{\ell m \ell' m'}$ of each block component $\mathbf{C}^{XY}$ are expressed as
\begin{equation}
    C^{XY}_{\ell m \ell' m'} = \frac{(4\pi)^2}{L^3} \sum_{\mathbf{n} \in N} \Delta^{X}_\ell(k_n) \Delta_{\ell'}^{Y*}(k_n) \xi_{k_n \ell m}^{X; \hat{\mathbf{k}}_{\mathbf{n}}} \xi_{k_n \ell' m'}^{Y; \hat{\mathbf{k}}_{\mathbf{n}} *} \frac{2\pi^2 \mathcal{P}(k_n)}{k_n^3},
\end{equation}

where $X, Y \in \{T, E, B\}$ denote temperature fluctuations, E-mode polarization, and B-mode polarization, respectively. For the scalar-induced $TT$, $TE$, and $EE$ blocks, we use the scalar power spectrum $\mathcal{P}=\mathcal{P}_{\mathcal{R}}(k_n)$ and the factor $\xi$ defined in Eq.~(\ref{scalar_xi}) for $T$, and Eq.~(\ref{tensor_xi}) for $E$. For the tensor-induced $BB$ block, we use the tensor power spectrum $\mathcal{P}=\mathcal{P}_{h}(k_n)$ and the factor $\xi$  defined in Eq.~(\ref{tensor_xi}). The transfer functions $\Delta_\ell^{X}$ with $X \in \{T, E, B\}$ correspond to the relevant temperature and polarization observables.

{\color{magenta}

\begin{figure}[htbp]
\centering
\includegraphics[width=\textwidth]{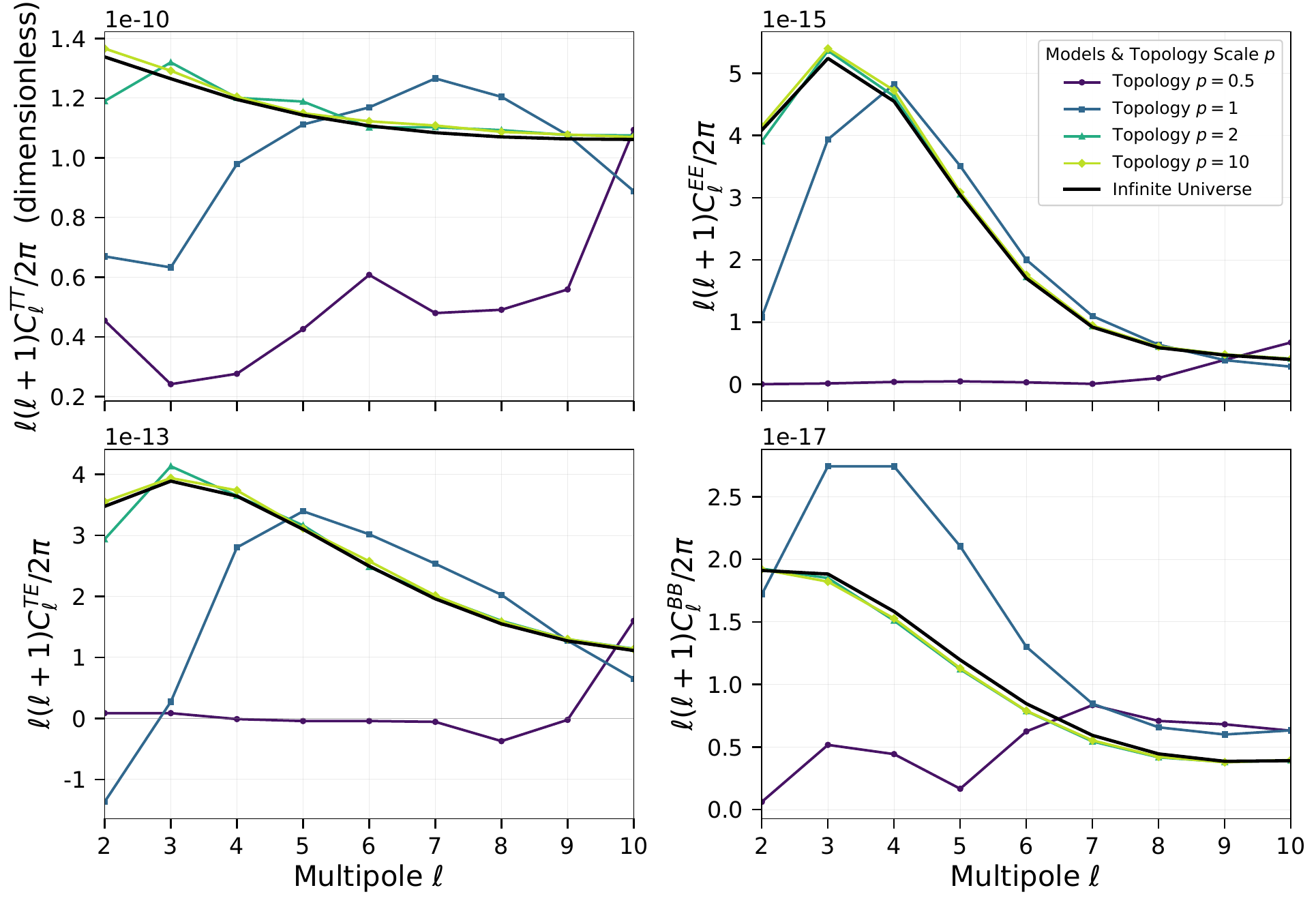}
\caption{ Comparison of the angular power spectra between the $\Lambda$CDM model
    with a trivial topology and the $E_1$ topology model.
    The top-left, top-right, bottom-left, and bottom-right panels display
    the TT, EE, TE, and BB power spectra, respectively.
    For the evaluation of the BB spectrum, a tensor-to-scalar ratio of
$r=0.01$ is assumed for both models.}

\label{fig:LambdaCDM_and_E1_Cl}

\end{figure}

}
As a check of the normalization and the diagonal angular power,
we compare the angular power spectra obtained from the $E_1$
covariance matrix with those computed using CAMB~\cite{2000ApJ...538..473L}.

\footnote{We use the CAMB normalization convention for tensor transfer functions. In this convention,
$C_\ell^{BB} = \left(\frac{\pi}{4}\right) \int_0^\infty \mathcal{P}_\mathcal{H}(k) \left[ \Delta_\ell^B(k) \right]^2 {\rm d}\ln k$, corresponding to an overall factor of $1/16$ relative to the scalar normalization $4\pi$~\cite{2004PhRvD..70d3011L}.
}
In Fig.~\ref{fig:LambdaCDM_and_E1_Cl}, we compare the angular power spectra between the models with a trivial topology and the $E_1$ topology.

For the comparison in Fig.~\ref{fig:LambdaCDM_and_E1_Cl}, we derive the
angular power spectra directly from the analytically computed covariance
matrix.  For a statistically anisotropic model, we define the angular
power spectrum as the rotationally averaged diagonal part of the
harmonic-space covariance,
\begin{equation}
    C_{\ell}^{XY}
    =
    \frac{1}{2\ell+1}
    \sum_{m=-\ell}^{\ell}
    C_{\ell m,\ell m}^{XY}.
\end{equation}
This definition reduces to the usual angular power spectrum in the
statistically isotropic limit, where
\begin{equation}
    C_{\ell m,\ell' m'}^{XY}
    \rightarrow
    C_{\ell}^{XY}
    \delta_{\ell\ell'}
    \delta_{mm'}
\end{equation}
as $L\rightarrow\infty$.
The off-diagonal components with $\ell\neq\ell'$ or $m\neq m'$ are not
represented in the one-dimensional angular power spectrum $C_\ell$;
instead, they contain additional information associated with statistical
anisotropy induced by the non-trivial topology.  These off-diagonal
components are retained in the full covariance matrix and are included
in the likelihood analysis below.

As shown in Fig.~\ref{fig:LambdaCDM_and_E1_Cl}, the power spectra converge toward the CAMB predictions as the size of the fundamental domain increases, i.e., the parameter $p$ increases. 
This convergence reflects the fact that the discrete wavevector sum approaches the continuous Fourier integral in the limit $L \to \infty$. For small fundamental domains, by contrast, the discreteness of the allowed modes and the absence of sufficiently long-wavelength modes produce substantial deviations, particularly at low multipoles.

\subsection{Numerical implementation}
\label{subsec:numerical}
In the $E_1$ topology model, the continuous Fourier integral is replaced by an infinite sum over discrete wavevectors. Numerical evaluation therefore requires truncating the mode sum at a finite maximum wavenumber. In this subsection, we describe the construction of the discrete wavevector grid, the choice of the cutoff scale, and the numerical implementation.

In the numerical calculation, we introduce a spherical cutoff \(k_{\rm max}\) in wavenumber space.
For the cubic \(E_1\) topology, the allowed wavevectors are discretized as
\begin{equation}
\mathbf{k}_{\mathbf n} = \frac{2\pi}{L}\mathbf n, \qquad \mathbf n=(n_x,n_y,n_z)\in\mathbb Z^3, 
\end{equation}
where \(L\) is the side length of the fundamental domain. In the present calculation, we include in the covariance sum all discrete modes satisfying
$$ |\mathbf{k}_{\mathbf n}|\le k_{\rm max}. $$
Equivalently, this corresponds to selecting all lattice points on the integer grid that satisfy
\begin{equation}
 n_x^2+n_y^2+n_z^2 \le \left(\frac{k_{\rm max}L}{2\pi}\right)^2. 
\end{equation}

The cutoff \(k_{\rm max}\) must be chosen to balance the required numerical accuracy against the computational cost. If \(k_{\rm max}\) is too small, the contribution from high-wavenumber modes is insufficiently included, leading to truncation errors in the covariance matrix. On the other hand, choosing \(k_{\rm max}\) unnecessarily large significantly increases the number of discrete modes included in the calculation and therefore the total computation time.

\begin{figure}[htbp]
\centering
\includegraphics[width=\textwidth]{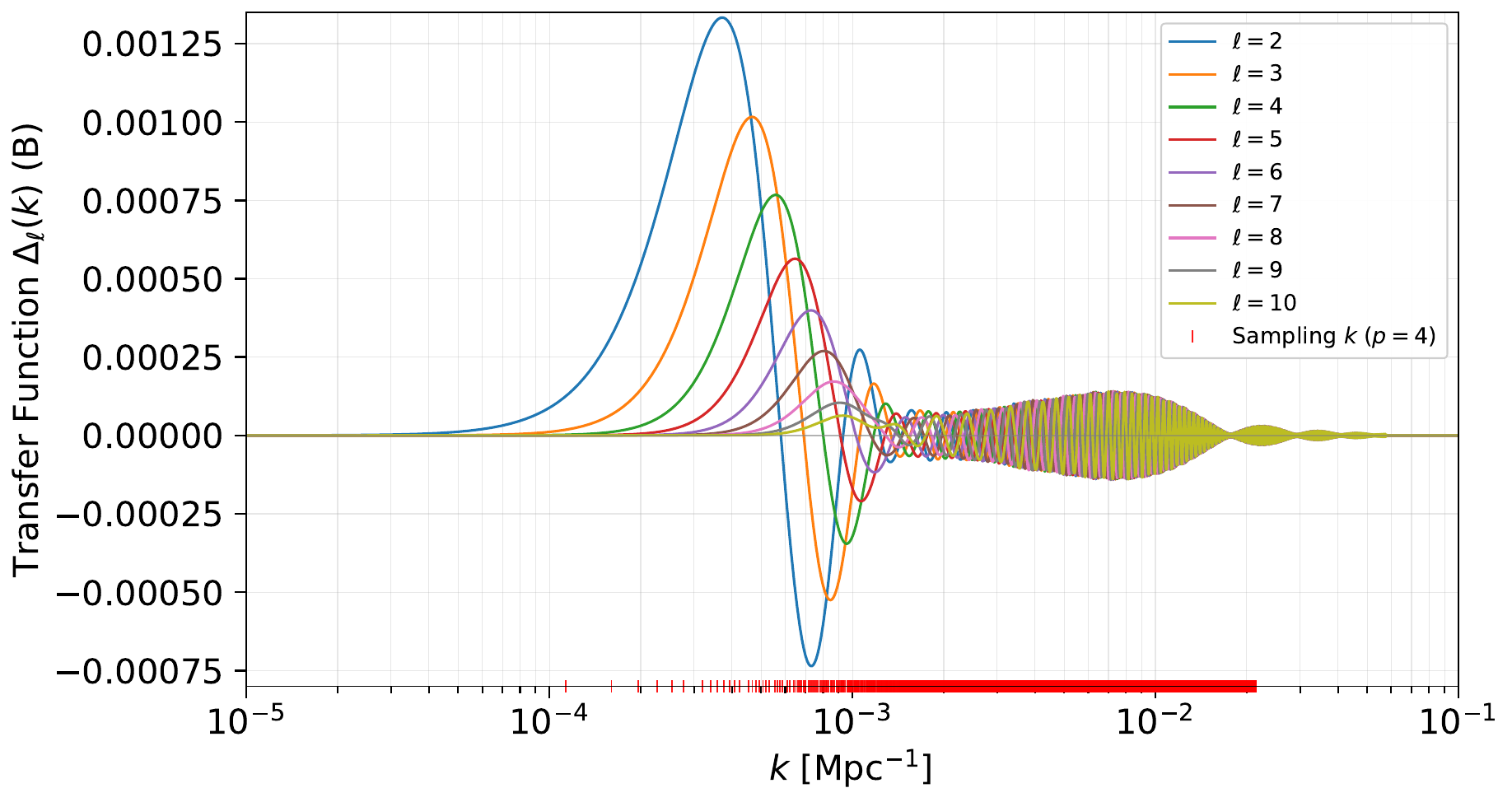}
\caption{B-mode transfer functions. 
As is evident from the figure, the transfer functions for all $\ell$ considered in this work sufficiently decay in the region $k \gtrsim 3 \times 10^{-2}$ Mpc$^{-1}$. Consequently, we adopted this value as the upper cutoff limit for the summation over the wavenumber space in our calculations. The red points on the bottom represent the wavenumbers sampled in the discrete sum to calculate the covariance matrix for the $E_1$ topology model with $p=4$. }
\label{Relationship between the B-mode transfer function and the maximum wavenumber sampling}

\end{figure}

In the \(E_1\) topology, continuous rotational symmetry is broken, so the angular dependence cannot in general be reduced to a one-dimensional integral over \(k\) alone. Instead, the mode sum must be evaluated directly over the three-dimensional discrete wavevector lattice. The number of modes contained within the spherical cutoff grows approximately as
$$ N_{\rm mode} \propto \left(\frac{k_{\rm max}L}{2\pi}\right)^3, $$
and hence the computational cost also scales roughly as the cube of this dimensionless cutoff.

Next, we determine the specific value of \(k_{\rm max}\) used in the numerical calculation. In the continuous limit, the contribution of each wavenumber mode to angular power spectra can be evaluated as
\begin{equation}
 C_\ell^{XY} \propto I_\ell \equiv \int d\ln k\, {\cal P}(k)   \Delta_\ell^X(k)\Delta_\ell^Y(k), 
\end{equation}
where \({\cal P}(k)\) is the primordial power spectrum and \(\Delta_\ell^X(k)\) is the X-mode transfer function.
For each multipole \(\ell\), we determine the wavenumber \(k_{99}^{(\ell)}\) at which the cumulative contribution to this integral reaches \(99\%\) of the total contribution. To ensure that a sufficiently wide wavenumber range is included for all multipoles in the range \(2\leq\ell\leq\ell_{\rm max}\), we then define
\begin{equation}
 k_{99} = \max_{2\leq\ell\leq\ell_{\rm max}} k_{99}^{(\ell)}. 
\end{equation}
To provide an additional numerical margin, the final cutoff is chosen as
$$ k_{\rm max}=1.1\,k_{99}. $$

As an example, figure~\ref{Relationship between the B-mode transfer function and the maximum wavenumber sampling} shows the B-mode transfer functions for \(2\leq\ell\leq10\), together with the wavenumbers sampled by the discrete modes. The transfer functions rapidly decay toward higher wavenumbers, consistent with the value of \(k_{\rm max}\) determined from the cumulative-contribution criterion described above.

As a convergence check, we increased the cutoff from
$k_{\rm max}=0.02144\,{\rm Mpc}^{-1}$ to
$0.022\,{\rm Mpc}^{-1}$ for all values of $p$ used in the likelihood analysis.
The resulting covariance matrices change by at most $3.3\times10^{-3}\%$
in relative Frobenius norm. The off-diagonal parts change by at most
$0.0081\%$, and the rotationally averaged diagonal power $C_\ell$ changes
by less than $0.087\%$ for all $2\leq\ell\leq10$

\section{Method and Result}
\label{sec:method}
\subsection{Gaussian likelihood}
Firstly,  we introduce a parameter $p$ to describe the topology scale as 
\begin{equation}
    L = p d_{\rm LSS}~,
\end{equation} 
where $d_{\rm LSS}$ is the comoving distance to the last-scattering surface.
We quantify the statistical distinguishability between the standard spatially flat $\Lambda$CDM model with trivial topology and the cubic $E_1$ topology using a likelihood-based analysis. Because the $E_1$ topology induces correlations between different spherical-harmonic modes, the relevant information is not fully captured by the conventional angular power spectra. We therefore arrange the harmonic coefficients of the observables into a data vector $\bm{a}$ and characterize each model by the corresponding harmonic-space covariance matrix, $\bm{C}_{\rm model} = \left< \bm{a} \bm{a}^\dagger\right>$.

Assuming that the harmonic coefficients follow a multivariate complex Gaussian
distribution with the reality condition, the log-likelihood for a given model covariance
$\mathbf{C}_{\rm model}$ is written as
\begin{equation}
    \label{eq:loglikelihood}
    \ln \mathcal{L}(\mathbf{a} | \mathbf{C}_{\rm model})
    =
    -\frac{1}{2} \mathbf{a}^\dagger
    \mathbf{C}_{\rm model}^{-1} \mathbf{a}
    - \frac{1}{2} \ln \det \mathbf{C}_{\rm model}
    - \frac{N}{2} \ln(2\pi),
\end{equation}
where $N$ is the number of independent real degrees of freedom, namely, $N=n_X\times \sum_{\ell_{\rm min}}^{\ell_{\rm max}} (2\ell+1)$, where the factor $n_X=3$ corresponds to the ($T+E+B$) analysis.
A larger value of
$\ln \mathcal{L}$ indicates that the assumed covariance matrix provides a better
statistical description of the data.

\subsection{Mock-data generation and analysis setup}
In this work, we generate mock CMB data assuming the fiducial 
$\Lambda$CDM model with trivial topology. We then evaluate
the likelihood in Eq.~\eqref{eq:loglikelihood} using the covariance matrices predicted by the $E_1$ topology models with different size parameters $p$. Although the $E_1$ covariance matrices are computed directly from the
analytical expression rather than estimated from mock realizations, the
expression contains an infinite sum over discrete wave modes. In the
numerical calculation, this sum is truncated at a finite $k_{\rm max}$,
and we therefore check the convergence of the covariance matrix with
respect to this cutoff.
By comparing these likelihood values with that obtained for the fiducial $\Lambda$CDM covariance matrix, we evaluate the likelihood-ratio statistic (equivalently $\Delta \chi^2$) between each $E_1$ topology model and the fiducial model.

Our aim is not to forecast the performance of a particular experiment, but to compare the intrinsic amount of topological information carried by the temperature, E-mode, and primordial B-mode polarization anisotropies. We therefore adopt an ideal cosmic-variance-limited setup and neglect instrumental noise, foreground residuals, and gravitational lensing. This provides a clean benchmark for assessing the maximum additional information that primordial B-mode polarization can contribute within the assumptions adopted here.

To compare consistently the information provided by the different data combinations, namely, temperature only ($T$), temperature plus $E$-mode polarization ($T+E$), and the full CMB ($T+E+B$), 

for each mock realization, we first generate a single complete fiducial \(\Lambda\)CDM data vector containing the T-, E-, and B-mode harmonic coefficients. The T, T+E, and T+E+B data vectors are then constructed by selecting the corresponding components from this same realization. Thus, the temperature coefficients \(a_{\ell m}^{T}\) are identical in all three analyses, and the E-mode coefficients \(a_{\ell m}^{E}\) are identical in the T+E and T+E+B analyses.

We generate mock data from the fiducial $\Lambda$CDM model with trivial
topology. Since the harmonic coefficients of a real CMB field satisfy
$a_{\ell,-m}^{X}=(-1)^m(a_{\ell m}^{X})^*$, the simulations are generated
using the corresponding independent real degrees of freedom. Their covariance
matrix in real space, $\boldsymbol{\Sigma}_{\Lambda{\rm CDM}}$, is obtained from the
harmonic-space covariance matrix
$\mathbf{C}_{\Lambda{\rm CDM}}$ through the unitary transformation described
in Appendix~\ref{app:real_representation}. We perform the Cholesky
decomposition
\begin{equation}
 \boldsymbol{\Sigma}_{\Lambda{\rm CDM}}
 =
 \mathbf{L}\mathbf{L}^{\mathsf T}
\end{equation}
and generate
\begin{equation}
 \mathbf{x}_{\rm mock}
 =
 \mathbf{L}\mathbf{z},
 \qquad
 \mathbf{z}\sim\mathcal{N}(\mathbf{0},\mathbf{I}).
\end{equation}
The variables $\mathbf{x}_{\rm mock}$ are then transformed back to the
complex harmonic coefficients $\mathbf{a}_{\rm mock}$. This procedure
guarantees that each realization satisfies the reality condition and that
$\langle\mathbf{a}_{\rm mock}\mathbf{a}_{\rm mock}^{\dagger}\rangle
=\mathbf{C}_{\Lambda{\rm CDM}}$.

We generate 100 independent mock realizations from the fiducial
$\Lambda$CDM covariance matrix. For each realization, we evaluate the
log-likelihood in Eq.~\eqref{eq:loglikelihood} using both the fiducial
$\Lambda$CDM covariance matrix and the covariance matrices predicted by the
$E_1$ topology models with $p=2,\,2.1,\,2.2,\,2.5,\,3,$ and $4$.

We define the
likelihood-ratio statistic as
\begin{equation}
 \Delta\chi^2(p)
 \equiv
 -2\ln\frac{\mathcal{L}_{E_1}(p)}
                 {\mathcal{L}_{\Lambda{\rm CDM}}}
 =
 -2\left[
 \ln\mathcal{L}_{E_1}(p)
 -
 \ln\mathcal{L}_{\Lambda{\rm CDM}}
 \right].
 \label{eq:delta_chi2}
\end{equation}
A positive value of $\Delta\chi^2$ indicates that the fiducial
$\Lambda$CDM model provides a better fit to the mock realization than the
corresponding $E_1$ model. For each value of $p$, we report the mean of
$\Delta\chi^2$ over the 100 realizations, with error bars showing the standard
deviation of its realization-to-realization distribution.

As shown in previous studies \cite{2024JCAP...03..036E,2025JCAP...08..015S}, the signatures of cosmic topology are most prominent in the low-$\ell$ regime (large scales). Therefore, we restrict our analysis to the multipole range of $\ell = 2$ to $10$.

\subsection{Illustrative CMB realizations}
\label{subsec:illustrative_maps}
\begin{figure}[htbp]
\centering
\includegraphics[width=\textwidth]{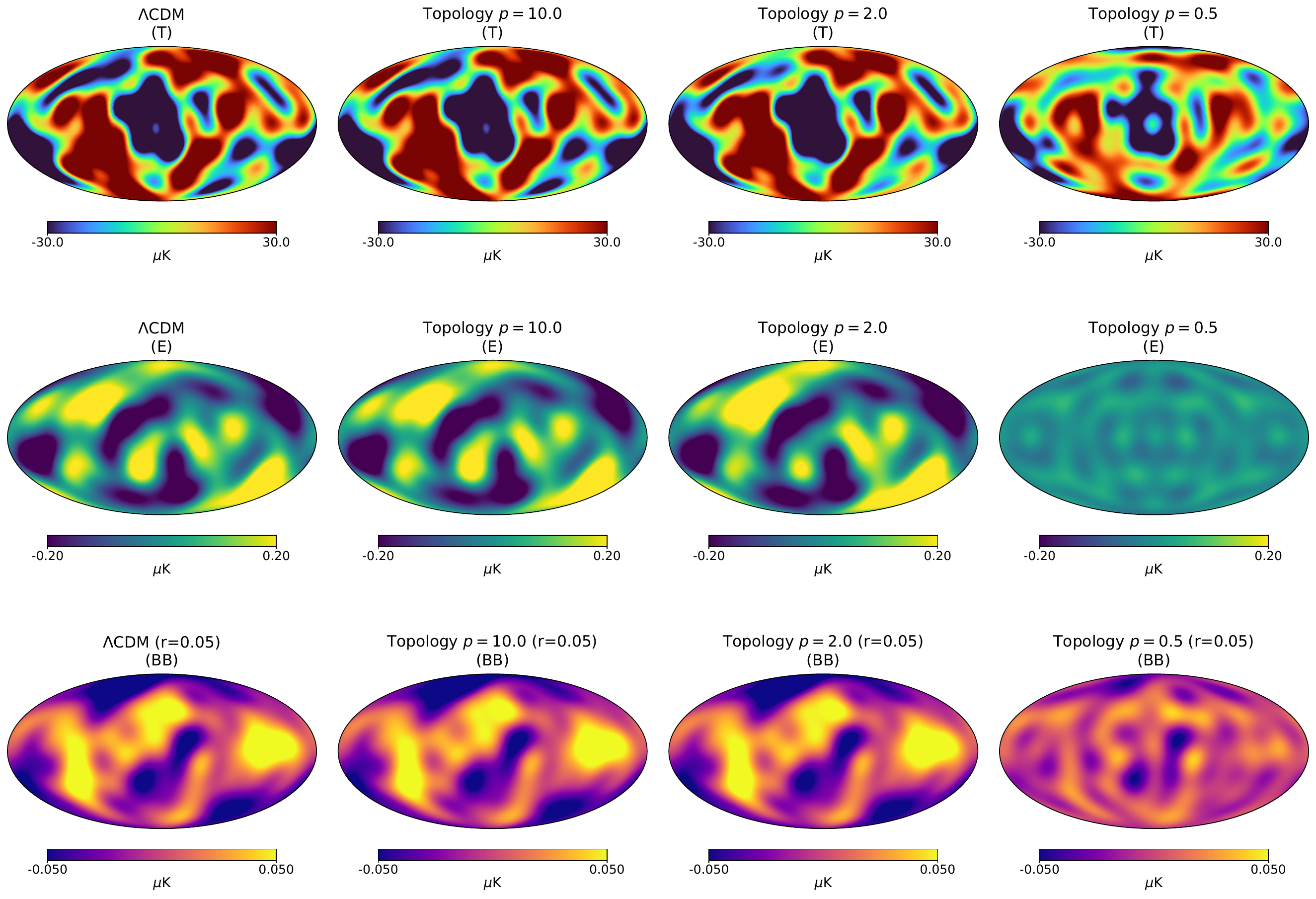}
\caption{Comparison between the standard $\Lambda$CDM model with trivial topology and cosmic topology models with different scale parameters ($p=10, 2$, and 0.5). The columns, from left to right, correspond to the $\Lambda$CDM model with trivial topology and the topology models with $p=10, p=2$, and $p=0.5$, respectively. The rows, from top to bottom, represent the temperature fluctuations (temperature fluctuations), E-mode polarization, and B-mode polarization components. To ensure consistent conditions for comparison, we fix the tensor-to-scalar ratio at $r=0.01$ across all models. Furthermore, to visualize the pure theoretical differences, we keep the initial random seed used to generate the spherical harmonic expansion coefficients identical for all models.\label{CMB_map}}
\end{figure}

To provide an intuitive illustration of the simulated data, we generate 
CMB maps from the harmonic coefficients used in the likelihood analysis using the HEALPix\footnote{\url{https://healpix.sourceforge.io}} package~\cite{2005ApJ...622..759G}.
Figure~\ref{CMB_map} shows representative realizations for the fiducial $\Lambda$CDM model with trivial topology and for the $E_1$ models with several values of $p$. All maps are truncated at $\ell_{\max}=10$.

Visual inspection shows that although the detailed phase patterns differ among realizations, the overall morphology remains qualitatively similar for $p=10$ and $p=2$.
The map for $p=0.5$, corresponding to a fundamental domain substantially smaller than the diameter of the last-scattering surface, displays a more visibly distinct large-scale morphology. Nevertheless, these examples should not be interpreted as a statistical comparison, because the appearance of an individual realization is strongly affected by cosmic variance.
Instead, we perform a quantitative comparison between the models using the covariance-based likelihood analysis introduced above.

\begin{figure}[htbp]
\centering
\includegraphics[width=\textwidth]{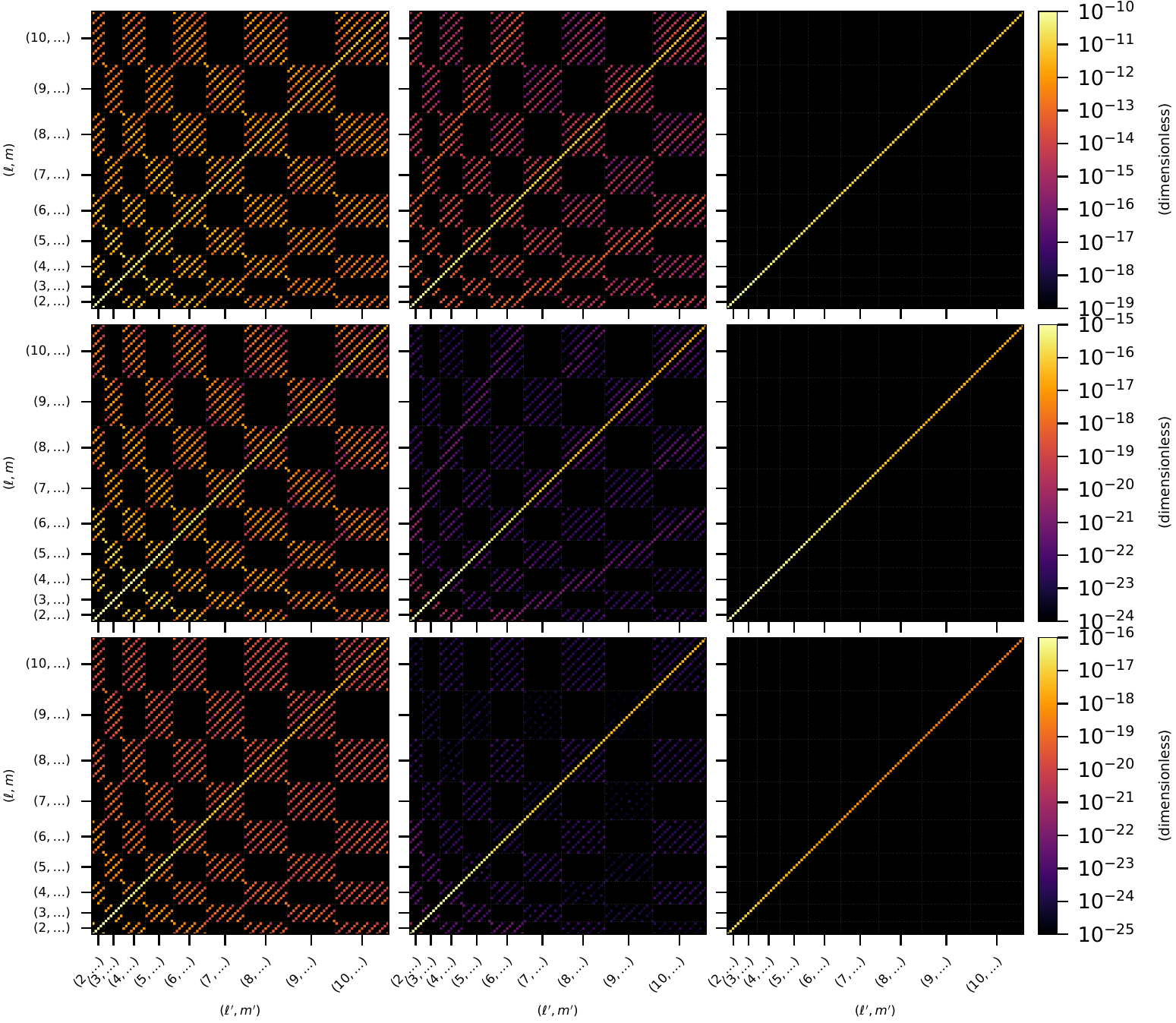}
\caption{
Absolute values of the CMB covariance matrices for the $E_1$ topology and
the fiducial $\Lambda$CDM model with trivial topology. The top, middle, and
bottom rows show the $TT$, $EE$, and $BB$ covariances, respectively. The
left and middle columns correspond to $p=2$ and $p=10$, while the right
column shows the fiducial model. The harmonic-mode pairs $(\ell,m)$ and
$(\ell',m')$ are represented by flattened one-dimensional indices on the
two axes.\label{fig:covariance_matrix}}
\end{figure}

\subsection{Covariance matrix for E$_1$ model}
\label{subsec:result_cov}

Figure~\ref{fig:covariance_matrix} shows the $TT$, $EE$, and $BB$ covariance matrices for the $E_1$ topology with $p=2$ and $p=10$, together with those of the fiducial $\Lambda$CDM model with trivial topology. 
For the trivial topology model, the covariance matrices are diagonal in $(\ell,m)$. 
By contrast, the $E_1$ topology induces off-diagonal correlations because the cubic fundamental domain breaks continuous rotational symmetry. These correlations are clearly visible for $p=2$ (left panels). 
Their amplitudes decrease as the parameter $p$ increases, and the covariance matrices for $p=10$ (middle panels) approach those of the fiducial $\Lambda$CDM model.

The B-mode covariance is also shown in the bottom panels. 
Although the latest observational results from BICEP/Keck~\cite{2021PhRvL.127o1301A} constrain the upper limit of the tensor-to-scalar ratio to $r < 0.036$, this figure assumes $r=0.05$  to emphasize the B-mode contribution visually. 
In the cosmic-
variance-limited analysis considered here, however, the likelihood-ratio statistic associated with the
primordial BB covariance is independent of the overall value of $r$, because the common
amplitude scaling cancels between the mock realizations and the model covariance.

The $BB$ covariance exhibits the same qualitative trend as the $TT$ and $EE$ covariances: topology-induced off-diagonal correlations are present for small fundamental domains and gradually disappear as $p$ increases; see Fig.~\ref{fig:covariance_matrix}.
However, their amplitudes are systematically smaller than those in the temperature and E-mode covariance matrices, especially for $p=2$. 
This already suggests that although primordial B-mode polarization carries information about the topology, its contribution to the overall discriminative power is expected to be limited. 
This difference reflects both the wavenumber ranges to which the corresponding transfer functions are sensitive and the resulting anisotropic correlations in harmonic space.We will discuss this point in more detail in Sec.~\ref{sec:discussion}.

\begin{figure}[htbp]
\centering
\includegraphics[width=\textwidth]{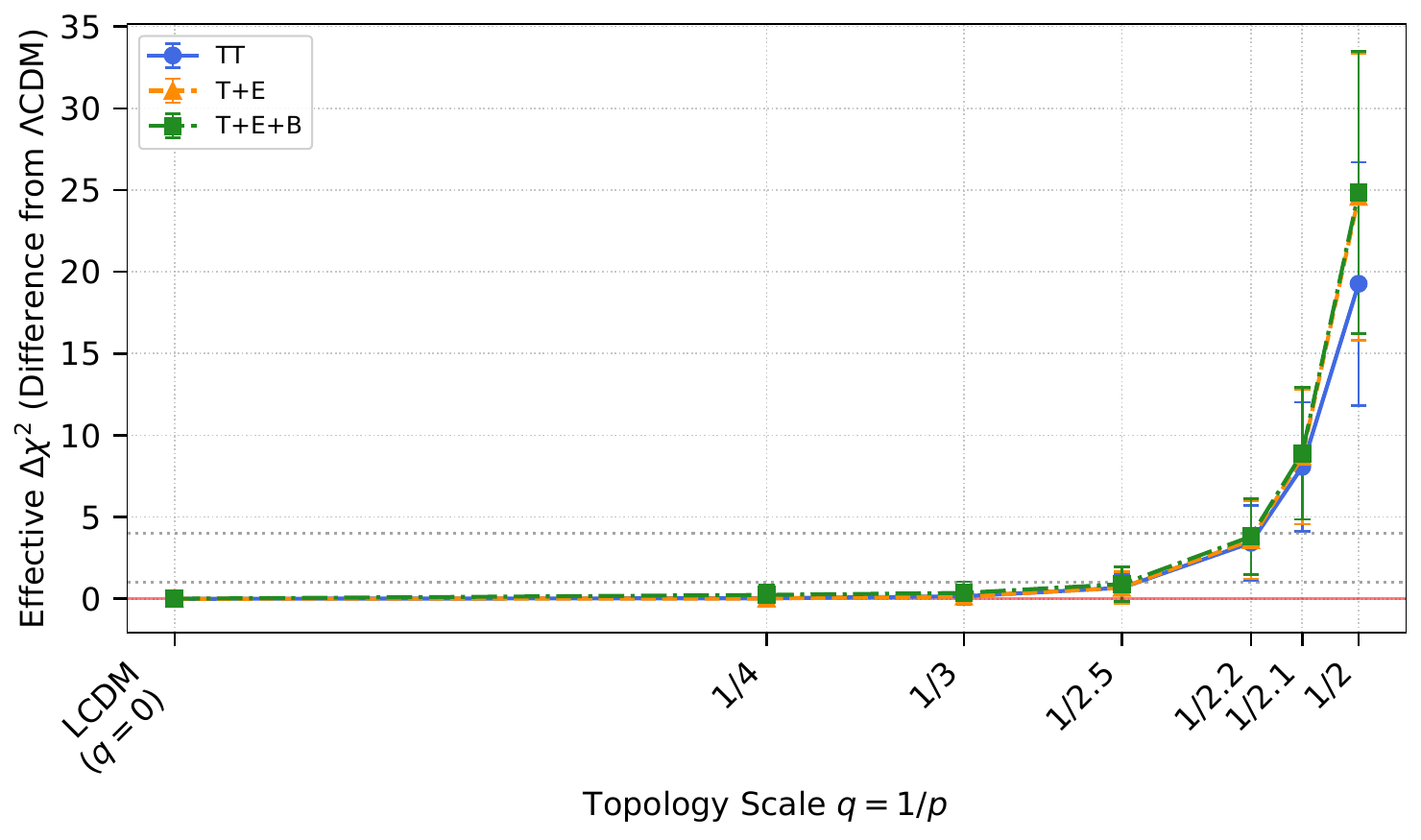}
\caption{Mean likelihood-ratio statistic $\Delta\chi^2$ obtained from 100 mock
realizations generated from the fiducial $\Lambda$CDM model with trivial
topology. Results are shown for the $T$, $T+E$, and $T+E+B$ data
combinations as a function of the inverse topology scale, $1/p$. The error
bars indicate the standard deviation of the realization-to-realization
distribution. The calculation assumes the cosmic-variance-limited limit
and neglects instrumental noise and gravitational lensing. A fiducial value
$r=0.05$ is adopted for the primordial B mode; in this ideal limit, the
likelihood-ratio statistic is independent of the overall value of $r$.
}
\label{fig:main_result}
\end{figure}

\subsection{Likelihood-ratio analysis}
\label{subsec:likelihood_results}
Figure~\ref{fig:main_result} shows the mean likelihood-ratio statistic,
$\Delta\chi^2$, obtained from 100 mock realizations for the $T$, $T+E$,
and $T+E+B$ data combinations. We assume a
cosmic-variance-limited case, neglecting instrumental noise or gravitational
lensing effect. We adopt $r=0.05$ as a fiducial value for the numerical calculation, although the resulting likelihood ratio is independent of $r$ under the assumptions considered here.

The discriminating power decreases rapidly as the size of the fundamental
domain increases. For $p=2$, the $E_1$ model is clearly distinguishable
from the fiducial $\Lambda$CDM model. Adding E-mode polarization increases
$\Delta\chi^2$ relative to the temperature-only analysis.

To separate the contributions from the EE auto-covariance and the TE correlation, we compare three cases: (i) temperature only, (ii) $T+E$ with the TE covariance set to zero in both the $E_1$ and fiducial $\Lambda$CDM models, and (iii) the full $T+E$ covariance including the TE correlation. This comparison allows us to quantify how much of the improvement obtained by adding E-mode polarization arises from the EE auto-covariance alone and how much is additionally provided by its correlation with temperature. We define the EE contribution as the increase in the expected $\Delta\chi^2$ from the temperature-only case to the $T+E$ case with $TE=0$, and the TE contribution as the further increase obtained by restoring the TE covariance. For $p=2$, the EE contribution is $2.25$, whereas the TE contribution is $8.95$, indicating that the TE correlation accounts for about $80\%$ of the total increase in the expected $\Delta\chi^2$ obtained by adding E-mode polarization. At $p=2.1$, the corresponding EE and TE contributions are $0.82$ and $0.41$, respectively, with the EE contribution becoming larger than that from TE. For $p\gtrsim2.2$, both contributions become small in absolute magnitude, consistent with the rapid decrease of the overall topological signal as the fundamental-domain size increases.

Primordial B-mode polarization provides a smaller but nonzero additional
contribution, as also confirmed by a direct comparison of the $T$ and $T+B$
data combinations. Under the present assumption of vanishing TB and EB
correlations, this B-mode contribution is the same whether B modes are added
to $T$ alone or to the combined $T+E$ data.

The likelihood-ratio statistic decreases rapidly, rather than linearly, as
$p$ increases. By $p\simeq2.5$, the mean $\Delta\chi^2$ has already fallen
below unity, and it continues to approach the fiducial $\Lambda$CDM result
as the fundamental-domain size increases ($q=1/p\rightarrow0$). This shows
that once the topology scale becomes sufficiently larger than the observable
region, distinguishing the $E_1$ model from the trivial-topology
$\Lambda$CDM model becomes increasingly difficult.

\section{Discussion}
\label{sec:discussion}
\subsection{Origin of the weak B-mode contribution}
The main result of this work is that adding primordial B-mode polarization
to the temperature and E-mode data produces only a marginal improvement in
the discrimination between the cubic $E_1$ topology and the fiducial
$\Lambda$CDM model with trivial topology. Previous covariance-based studies
have demonstrated that non-trivial Euclidean topologies can be probed
through correlations between different spherical-harmonic modes, even when
the fundamental domain is sufficiently large that matched-circle signatures
are absent or difficult to detect
\cite{2016A&A...594A..18P,2024PhRvL.132q1501A,2024JCAP...03..036E}.
Our results extend this approach by quantifying the incremental information
provided by primordial B-mode polarization.

The likelihood analysis reveals a clear hierarchy in the sensitivity to
the $E_1$ topology, with the temperature anisotropy providing the strongest
constraint, followed by the E-mode and B-mode polarization. This hierarchy
can be understood from two closely related aspects of the topology-induced
covariance: the wavenumber dependence of the CMB transfer functions and the
resulting off-diagonal correlations in harmonic space.

\begin{figure}[htbp]
\centering
\includegraphics[width=\textwidth]{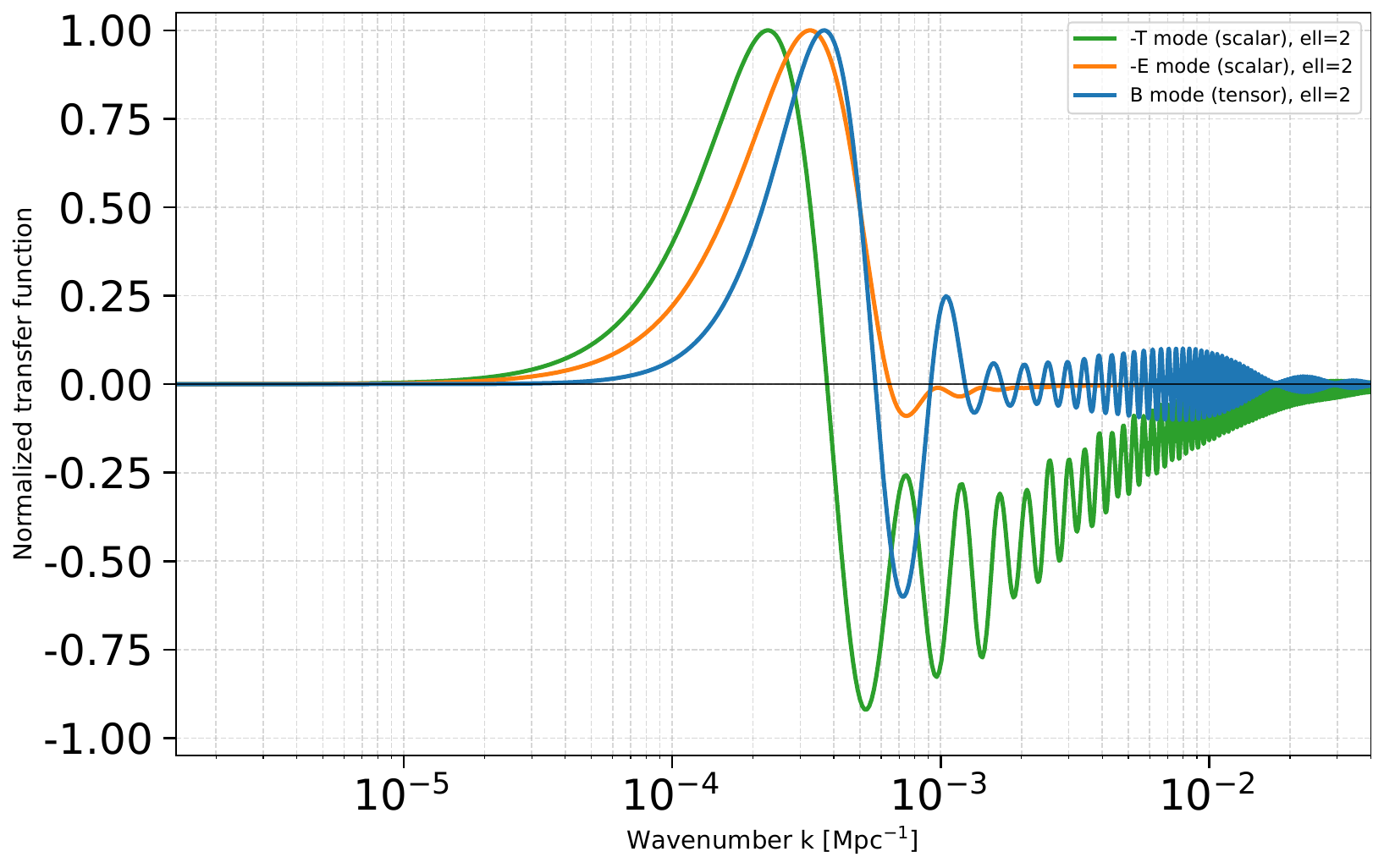}
\caption{
Normalized transfer functions of the CMB anisotropies at
$\ell=2$: temperature ($T$), E-mode polarization ($E$), and primordial
B-mode polarization ($B$). The onset of the response shifts from lower to
higher wavenumbers in the order $T \rightarrow E \rightarrow B$, showing
that the temperature anisotropy is sensitive to the largest spatial scales,
whereas the B mode responds only at comparatively larger $k$. This
difference helps explain the hierarchy of constraining power for the
$E_1$ topology found in the likelihood analysis.
}
\label{fig:normalized_transfer_functions}
\end{figure}

The compact topology discretizes the allowed Fourier modes and removes
modes with wavelengths larger than the fundamental domain. The impact of
this modification on a given CMB observable depends on its transfer
function, which determines how strongly different wavenumbers contribute
to the observed anisotropies. In particular, the low-$\ell$ temperature
transfer function has support at smaller wavenumbers than the E-mode
transfer function, while the B-mode response turns on at still larger
wavenumbers, as shown in Fig.~\ref{fig:normalized_transfer_functions}. The temperature anisotropy is therefore more sensitive to the
largest-scale modes, for which the effects of the compact topology are most
pronounced.

\begin{figure}[t]
    \centering
    \includegraphics[width=\textwidth]{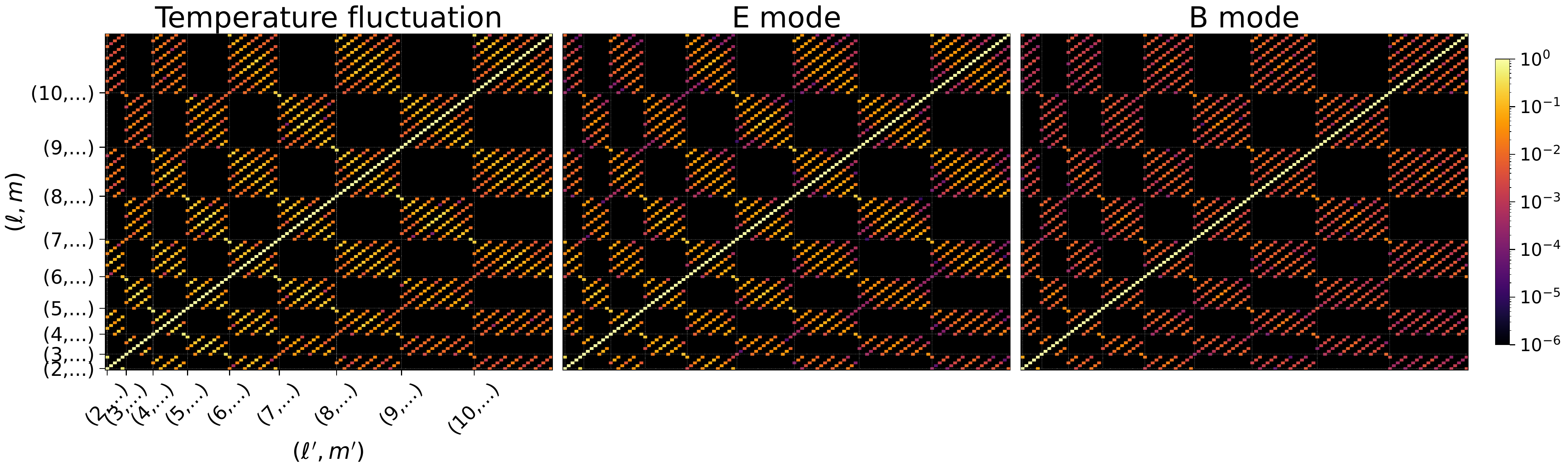}
    \caption{Correlation matrices corresponding to the $TT$ (left), $EE$ (middle), and $BB$ (right)
covariance matrices for the $E_1$ topology with $p=2$. Each element is normalized according to Eq.~(\ref{eq:normalize}) so that all diagonal elements are unity. This normalization removes the overall signal amplitude and highlights the relative strength of the topology-induced off-diagonal correlations. Compared with the temperature and E-mode polarization, the B-mode polarization exhibits substantially weaker topology-induced off-diagonal correlations. }
    \label{fig:covariance}
\end{figure}

At the same time, the topology breaks statistical isotropy and produces
correlations between different $(\ell,m)$ modes. These off-diagonal
correlations constitute an additional and characteristic signature of the
$E_1$ topology. 
To isolate the relative amplitude of the off-diagonal contributions, we normalize the covariance matrix as
\begin{equation}
R^{XX}_{\ell m,\ell' m'}
=
\frac{
 \left|C^{XX}_{\ell m,\ell' m'}\right|
}{
 \sqrt{
 \left|C^{XX}_{\ell m,\ell m}\right|
 \left|C^{XX}_{\ell' m',\ell' m'}\right|
 }
}.
\label{eq:normalize}
\end{equation}
As shown in Fig.~\ref{fig:covariance}, their relative amplitudes
also differ among the temperature, E-mode, and B-mode observables. 
The weaker off-diagonal structure of the B-mode covariance is therefore
consistent with its weaker sensitivity to the topology-sensitive
large-scale Fourier modes.

These two effects should not be regarded as independent. The transfer
functions determine how the discrete topological eigenmodes are projected
onto the CMB observables, and thereby also affect the resulting
off-diagonal covariance structure. The limited constraining power of the
B mode can thus be understood as the combined consequence of its weaker
response to the largest spatial scales and the correspondingly weaker
topology-induced off-diagonal correlations in harmonic space.

\subsection{Relation to previous studies}

It is useful to place our result in the context of previous studies of tensor-induced signatures of cosmic topology. Ref.~\citep{2025JCAP...08..015S} investigated the topological information contained in tensor-induced CMB anisotropies and showed that their Kullback–Leibler divergence can continue to increase as higher multipoles are included. Our analysis addresses a complementary but distinct question: how much additional constraining power primordial B-mode polarization provides once the scalar-dominated temperature and E-mode information is already included. We find that, over the low-multipole range considered here, the incremental improvement from adding the $BB$ covariance is small.
The difference also reflects the approximations adopted in the present analysis. We treat the temperature and E-mode anisotropies as purely scalar-induced and the B-mode anisotropy as purely tensor-induced, so that the covariance is block diagonal and the $TB$ and $EB$ blocks vanish. By contrast, a full tensor calculation in a non-trivial topology generally includes tensor contributions to the $TT$, $TE$, and $EE$ blocks, as well as topology-induced $TB$ and $EB$ correlations \citep{2025JCAP...08..015S}. Our result should therefore be interpreted specifically as a quantification of the additional information carried by the primordial $BB$ block beyond the scalar $T+E$ information. Extending the analysis to the full tensor covariance and to higher multipoles will be necessary to assess the total constraining power of tensor perturbations on cosmic topology.

\subsection{Future Prospects}
The present analysis intentionally adopts an idealized cosmic-variance-limited
setup to isolate the intrinsic topological information contained
in the CMB covariance. A realistic experimental forecast would additionally
require instrumental noise, gravitational-lensing B-modes, Galactic
foreground residuals, incomplete sky coverage, and instrumental
systematics. Since the primordial B-mode already provides only a 
marginal
incremental improvement in the ideal limit considered here, including these
observational effects is not expected to change the qualitative conclusion
that temperature and E-mode polarization provide the dominant sensitivity
within the low-multipole range studied in this work.

A more important extension is to investigate how the relative contributions
of the diagonal power and the topology-induced off-diagonal correlations
change with multipole range. Extending the analysis to higher multipoles
and including the full tensor-induced covariance, including tensor
contributions to $TT$, $TE$, and $EE$ as well as $TB$ and $EB$
correlations \cite{2025JCAP...08..015S}, will be necessary for assessing the ultimate information
content of tensor perturbations.

For the Gaussian and linear framework considered in this work, the full
harmonic-space covariance completely specifies the statistical information
contained in the CMB maps. Machine-learning methods therefore cannot
extract additional information beyond that contained in the exact Gaussian
likelihood. They may nevertheless offer computational advantages when
extending topology searches to higher multipoles, multiple topology classes
and orientations, or realistic observational effects, for which direct
evaluation of the full covariance likelihood becomes increasingly
expensive. Indeed, machine-learning approaches have recently been explored
for the classification of cosmic topology and likelihood-free inference
\cite{2024JCAP...09..057T}.

The situation may be different if extensions of the present framework
produce genuinely non-Gaussian signatures that are not fully characterized
by the two-point covariance. For example, non-Gaussian CMB statistics have
been found in studies of compact hyperbolic topologies even when the
primordial fluctuations are assumed to be Gaussian
\cite{PhysRevD.62.103001}. In such cases, methods capable of extracting
complex higher-order features directly from sky maps, including
machine-learning approaches, could provide information that is not captured
by a Gaussian covariance likelihood. Exploring whether analogous
non-Gaussian signatures arise in more general topology models or through
realistic nonlinear and observational effects would therefore be an
interesting direction for future work.

\section{Conclusion}
In this work, we investigated the additional information provided by
primordial CMB B-mode polarization for constraining the cubic $E_1$
cosmic topology. Using the full harmonic-space covariance matrices of the
temperature, E-mode, and B-mode anisotropies, we performed a likelihood
analysis based on mock realizations generated from a fiducial
$\Lambda$CDM model with trivial topology.

We deliberately considered the ideal cosmic-variance-limited case in
order to explore the intrinsic sensitivity of each CMB observable to
cosmic topology. Even under these idealized conditions, the inclusion of
primordial B-mode polarization produces only a 
marginal improvement over the
$T+E$ analysis. The hierarchy in the constraining power of the temperature,
E-mode, and B-mode anisotropies can be understood through the combined
effects of their transfer-function responses to the topology-sensitive
large-scale modes and the corresponding topology-induced off-diagonal
correlations in harmonic space.

Our results therefore indicate that primordial B-mode polarization carries
only limited additional information on the $E_1$ topology at the low
multipoles $\ell\leq 10$. The covariance-based analysis developed in
this work provides a useful idealized benchmark for future studies
including a wider multipole range, the full tensor covariance, and
realistic observational effects.

\appendix

\section{Real representation of the harmonic-space likelihood}
\label{app:real_representation}

The spherical-harmonic coefficients of a real CMB field satisfy the
reality condition
\begin{equation}
 a_{\ell,-m}^{X}
 =
 (-1)^m
 \left(a_{\ell m}^{X}\right)^*,
 \label{eq:reality_condition}
\end{equation}
where $X\in\{T,E,B\}$. Consequently, although it is convenient to collect
all coefficients with $-\ell\leq m\leq\ell$ into a complex data vector,
they contain only the same number of independent real degrees of freedom
as the corresponding real-space field.

In this appendix, we show that the Gaussian likelihood written in terms of
the full complex harmonic-space vector is equivalent to the ordinary
multivariate Gaussian likelihood for the independent real degrees of
freedom. We also describe how mock realizations satisfying
Eq.~\eqref{eq:reality_condition} can be generated from a theoretically
calculated harmonic-space covariance matrix.

For notational simplicity, we first consider a single CMB field and a fixed
multipole $\ell$. The extension to multiple multipoles and multiple fields is
obtained by applying the same transformation block by block.

For $m>0$, we define the independent real variables
\begin{equation}
 x_{\ell m}^{R}
 =
 \sqrt{2}\,\mathrm{Re}\,a_{\ell m},
 \qquad
 x_{\ell m}^{I}
 =
 \sqrt{2}\,\mathrm{Im}\,a_{\ell m},
 \label{eq:real_variables}
\end{equation}
together with
\begin{equation}
 x_{\ell 0}=a_{\ell 0}.
\end{equation}
The inverse relations are
\begin{equation}
 a_{\ell m}
 =
 \frac{x_{\ell m}^{R}+i x_{\ell m}^{I}}{\sqrt{2}},
 \qquad
 a_{\ell,-m}
 =
 (-1)^m
 \frac{x_{\ell m}^{R}-i x_{\ell m}^{I}}{\sqrt{2}}.
 \label{eq:inverse_real_variables}
\end{equation}

Introducing $s_m\equiv(-1)^m$, Eq.~\eqref{eq:inverse_real_variables} can be
written as
\begin{equation}
 \begin{pmatrix}
  a_{\ell m}\\
  a_{\ell,-m}
 \end{pmatrix}
 =
 U_m
 \begin{pmatrix}
  x_{\ell m}^{R}\\
  x_{\ell m}^{I}
 \end{pmatrix},
 \qquad
 U_m
 =
 \frac{1}{\sqrt{2}}
 \begin{pmatrix}
  1 & i\\
  s_m & -i s_m
 \end{pmatrix}.
 \label{eq:unitary_block}
\end{equation}
The matrix $U_m$ is unitary,
\begin{equation}
 U_m^\dagger U_m
 =
 U_m U_m^\dagger
 =
 I_2.
\end{equation}

By combining the $m=0$ component with the blocks in
Eq.~\eqref{eq:unitary_block}, and by applying the same construction to all
multipoles and CMB fields, we obtain
\begin{equation}
 \mathbf{a}=U\mathbf{x},
 \qquad
 \mathbf{x}=U^\dagger\mathbf{a},
 \label{eq:a_Ux}
\end{equation}
where $\mathbf{a}$ contains all harmonic coefficients, including both
positive and negative $m$, and $\mathbf{x}$ contains only independent real
degrees of freedom. The full matrix $U$ is unitary,
\begin{equation}
 U^\dagger U=UU^\dagger=I.
\end{equation}
Although $U^\dagger\mathbf{a}$ is formally a complex-linear transformation,
the reality condition in Eq.~\eqref{eq:reality_condition} guarantees that
the resulting vector $\mathbf{x}$ is real.

We define the harmonic-space covariance matrix by
\begin{equation}
 C
 \equiv
 \left\langle
  \mathbf{a}\mathbf{a}^\dagger
 \right\rangle,
 \label{eq:complex_covariance_appendix}
\end{equation}
and the covariance matrix of the independent real variables by
\begin{equation}
 \Sigma
 \equiv
 \left\langle
  \mathbf{x}\mathbf{x}^{\mathsf T}
 \right\rangle.
 \label{eq:real_covariance_appendix}
\end{equation}
Using Eq.~\eqref{eq:a_Ux}, these matrices are related by
\begin{equation}
 C
 =
 U\Sigma U^\dagger,
 \qquad
 \Sigma
 =
 U^\dagger C U.
 \label{eq:covariance_transformation}
\end{equation}
If the theoretical covariance matrix $C$ is consistent with the reality
condition, $\Sigma$ obtained from Eq.~\eqref{eq:covariance_transformation}
is real and symmetric.

The inverse covariance matrices satisfy
\begin{equation}
 C^{-1}
 =
 U\Sigma^{-1}U^\dagger.
\end{equation}
It follows that the quadratic forms in the complex and real
representations are identical:
\begin{align}
 \mathbf{a}^\dagger C^{-1}\mathbf{a}
 &=
 (U\mathbf{x})^\dagger
 U\Sigma^{-1}U^\dagger
 (U\mathbf{x})
 \nonumber\\
 &=
 \mathbf{x}^{\mathsf T}
 \Sigma^{-1}
 \mathbf{x}.
 \label{eq:quadratic_form_equivalence}
\end{align}

The determinants are also identical. From
Eq.~\eqref{eq:covariance_transformation},
\begin{align}
 \det C
 &=
 \det(U\Sigma U^\dagger)
 \nonumber\\
 &=
 \det U\,
 \det\Sigma\,
 \det U^\dagger.
\end{align}
Since $U$ is unitary,
\begin{equation}
 \det U\,\det U^\dagger
 =
 \det(UU^\dagger)
 =
 1,
\end{equation}
and hence
\begin{equation}
 \det C=\det\Sigma.
 \label{eq:determinant_equivalence}
\end{equation}

The ordinary multivariate Gaussian likelihood for the independent real
variables is
\begin{equation}
 \ln\mathcal{L}
 =
 -\frac{1}{2}
 \mathbf{x}^{\mathsf T}
 \Sigma^{-1}
 \mathbf{x}
 -\frac{1}{2}
 \ln\det\Sigma
 -\frac{N}{2}\ln(2\pi),
 \label{eq:real_gaussian_likelihood}
\end{equation}
where $N$ is the number of independent real degrees of freedom. Using
Eqs.~\eqref{eq:quadratic_form_equivalence} and
\eqref{eq:determinant_equivalence}, this can equivalently be written as
\begin{equation}
 \ln\mathcal{L}
 =
 -\frac{1}{2}
 \mathbf{a}^\dagger
 C^{-1}
 \mathbf{a}
 -\frac{1}{2}
 \ln\det C
 -\frac{N}{2}\ln(2\pi).
 \label{eq:complex_form_real_likelihood}
\end{equation}
Thus, Eq.~\eqref{eq:complex_form_real_likelihood} should be interpreted as
the real Gaussian likelihood expressed compactly in terms of the full complex
harmonic-space vector. It is not the likelihood of an unconstrained proper
complex Gaussian vector, for which the normalization and the factors of
$1/2$ would differ.

The same transformation provides a convenient procedure for generating
mock realizations. Given a theoretical harmonic-space covariance matrix
$C$, we first construct
\begin{equation}
 \Sigma=U^\dagger C U.
 \label{eq:sigma_from_C}
\end{equation}
We then perform the Cholesky decomposition
\begin{equation}
 \Sigma=LL^{\mathsf T}
\end{equation}
and generate
\begin{equation}
 \mathbf{x}_{\rm mock}
 =
 L\mathbf{z},
 \qquad
 \mathbf{z}
 \sim
 \mathcal{N}(\mathbf{0},I).
 \label{eq:real_mock_generation}
\end{equation}
Finally, the complex harmonic coefficients are obtained from
\begin{equation}
 \mathbf{a}_{\rm mock}
 =
 U\mathbf{x}_{\rm mock}.
 \label{eq:complex_mock_generation}
\end{equation}
Each realization generated in this way satisfies the reality condition, and
its ensemble covariance is
\begin{align}
 \left\langle
  \mathbf{a}_{\rm mock}
  \mathbf{a}_{\rm mock}^\dagger
 \right\rangle
 &=
 U
 \left\langle
  \mathbf{x}_{\rm mock}
  \mathbf{x}_{\rm mock}^{\mathsf T}
 \right\rangle
 U^\dagger
 \nonumber\\
 &=
 U\Sigma U^\dagger
 =
 C.
\end{align}

For multiple CMB fields, such as $T$, $E$, and $B$, the transformation $U$
acts independently on the $m$ indices of each field. Cross-covariances such
as $TE$ are retained in $\Sigma$, and the equivalence of the two likelihood
representations remains unchanged.

\acknowledgments
The numerical calculations in this study were performed using the Hubble computing cluster at Nagoya University. The authors acknowledge the use of the CAMB package for the analysis. Some of the results in this paper have been derived using the HEALPix package~\cite{2005ApJ...622..759G}. The authors express deep gratitude to Tetsuya Shiromizu at the Graduate School of Mathematics, Nagoya University, for fruitful discussions. A.J.N. is supported by JSPS KAKENHI Grant Number JP25H01551, JP25K01044, JP23H00108, and JP22K21349. K.I. is supported by JSPS KAKENHI Grant Number 24K00625 and the JST FOREST Program JPMJFR20352935.

\bibliographystyle{JHEP}
\bibliography{biblio.bib}

\end{document}